\documentclass[aps,prl,showpacs,nofootinbib, twocolumn]{revtex4-1}

\usepackage{amssymb}
\usepackage{amsmath}
\usepackage{bm}
\expandafter\ifx\csname package@font\endcsname\relax\else
\expandafter\expandafter
 \expandafter\usepackage
 \expandafter\expandafter
 \expandafter{\csname package@font\endcsname}%
\fi

\renewcommand{\prl}{{\it Phys. Rev. Lett.}}

\newcommand{\aj}{{\it Astron. J. (USA)}}

\begin{document}
\title{Carter-like Constant of Motion in Newtonian Gravity is the Vinti Integral}
\author{Sergei M. Kopeikin}
\email{kopeikins@missouri.edu}
\affiliation{Department of Physics \& Astronomy, University of
Missouri-Columbia, 65211, USA}
\begin{abstract}
\noindent We compare the Vinti integral of classic celestial mechanics with a conserved Carter-like integral of motion for an axially-symmetric body in the Newtonian theory that has been recently found by Clifford Will. We demonstrate that the integrals are identical. It sheds new light on the Newtonian limit of the Kerr geometry.
\end{abstract}
\maketitle\noindent
Recently, {\it Physical Review Letters} published a paper by C. ~M. Will \cite{will-car} that pointed out an "unusual" property of the Newtonian gravitational field of an axisymmetric body -- the existence of a "new integral of motion" that reproduces the relationship $Q_{2l}=(mQ_2/m)^l$ existing between the multipole moments of the Kerr black hole. The integral is given in the Cartesian coordinates ${\bm r}=(x,y,z)$ by \cite[equation 20]{will-car}
\begin{equation}
\label{1}
C=h^2+\frac{Q_2}{m}\left({\bm e}\cdot{\bm v}\right)^2-2m\hat z\sum^{\infty}_{l=0}\left(\frac{Q_2}{m}\right)^{l+1}\frac{P_{2l+1}(\hat z)}{r^{2l+1}}\;,
\end{equation}
where $r=|{\bm r}|=(x^2+y^2+z^2)^{1/2}$, $m$ is mass of the body, $Q_2$ is its quadrupole moment, ${\bm e}$ is the unit vector along the symmetry axis ($z$-axis), ${\bm n}={\bm r}/r$, $\hat z=:\left({\bm n}\cdot{\bm e}\right)=\cos\theta$, ${\bm h}={\bm r}\times{\bm v}$ is the total angular momentum of a test particle orbiting the body, $h=({\bm h}\cdot{\bm h})^{1/2}$, $P_l(\hat z)$ is the Legendre polynomial, and the universal gravitational constant $G=1$.

In fact, there is nothing "unusual" about integral (\ref{1}) as it is well-known in classic celestial mechanics as the Vinti integral \cite[equation 4.14a]{vinti}
\begin{equation}
\label{2}
C=2ER^2+2mR+\frac{b^2C^2_3}{R^2+b^2}-\frac{(R^2+b^2\cos^2\sigma)^2}{R^2+b^2}\dot R^2\;,
\end{equation}
where the overdot denotes a time derivative, $E$ is the integral of energy, $C_3=h_z={\bm e}\cdot{\bm h}$ is the conserved $z$-component of the angular momentum, $b$ is a free constant characterizing the central body, and $(R,\sigma,\phi)$ are the oblate spheroidal coordinates related to the Cartesian coordinates $(x,y,z)$ as follows
\begin{eqnarray}
\label{3}
r\sin\theta\cos\phi=&x&=\sqrt{R^2+b^2}\sin\sigma\cos\phi\;,\\
r\sin\theta\sin\phi=&y&=\sqrt{R^2+b^2}\sin\sigma\sin\phi\;,\\
r\cos\theta=&z&=R\cos\sigma\;.
\end{eqnarray}
Equivalence of two expressions, (\ref{1}) and (\ref{2}), is established by direct calculation after making identification $Q_2/m=-b^2$, and taking into account that:
\begin{itemize}
\item[(A)] the conserved energy $E=T+V$, where $T$ is the kinetic energy of the test particle in $(x,y)$ plane and $V$ is the effective potential defined by
\begin{eqnarray}
\label{4}
T&=&\frac12\frac{R^2+b^2\cos^2\sigma}{R^2+b^2}\dot R^2+\frac12(R^2+b^2\cos^2\sigma)\dot\sigma^2\;,\\
V&=&-\frac{mR}{R^2+b^2\cos^2\sigma}+\frac12\frac{C^2_3}{(R^2+b^2)\sin^2\sigma}\;,
\end{eqnarray}
\item[(B)] the very last term in equation (\ref{1}) can be re-written in the spheroidal coordinates as
\begin{equation}
\label{5}
\hat z\sum^{\infty}_{l=0}\left(\frac{Q_2}{m}\right)^{l+1}\frac{P_{2l+1}(\hat z)}{r^{2l+1}}
=-\frac{R b^2\cos^2\sigma}{R^2+b^2\cos^2\sigma}\;,
\end{equation}
\item[(C)] the total angular momentum of the particle
\begin{eqnarray}
h^2&=&2(R^2+b^2\sin^2\sigma)^2\left[T+\frac{1}{2\sin^2\sigma}\frac{C^2_3}{R^2+b^2}\right]\\\nonumber
&-&\left(R\dot R+b^2\sin\sigma\cos\sigma\dot\sigma\right)^2\;,
\end{eqnarray}
\item[(D)]  the $z$ component of the particle velocity
\begin{equation}
\left({\bm e}\cdot{\bm v}\right)=\dot z=\dot R\cos\sigma-R\sin\sigma\dot\sigma\;.
\end{equation}
\end{itemize}
The Vinti integral was discovered by J.~P. Vinti \cite{vin1,vin2} who showed that the problem of motion of the Earth satellite could be formulated in terms of a new geopotential having a special form where the Hamilton-Jacobi equation for the satellite is separable. Soviet scientists Aksenov, Grebenikov and Demin \cite{agd} independently re-discovered the Vinti potential and recognized that the Vinti problem is a transform of the Euler-Darboux problem \cite{vozm} of two fixed masses having both real and imaginary parts. Correspondence between the relativistic two-center problems and the Carter-like constant of motion has been studied by Mirshekari \& Will \cite{mirwill}. Enlightening theoretical treatment and interpretation of the Vinti integral is given in excellent textbooks by Math\'una \cite{vinti} and Vozmischeva \cite{vozm}.

\end{document}